\documentclass{acm_proc_article-sp}%
\usepackage{amsmath}
\usepackage{graphicx}%
\usepackage{amsfonts}%
\usepackage{amssymb}
\newtheorem{theorem}{Theorem}

\newtheorem{definition}[theorem]{Definition}

\newtheorem{lemma}[theorem]{Lemma}

\newtheorem{proposition}[theorem]{Proposition}

\begin{document}

\title{Lower Bounds for Local Search by Quantum Arguments}
\author{Scott Aaronson\thanks{University of California, Berkeley. \ Email:
\texttt{aaronson@cs.berkeley.edu}. \ Parts of this work were done at the
Hebrew University (Jerusalem, Israel) and the Perimeter Institute (Waterloo,
Canada). \ Supported by an NSF Graduate Fellowship, by NSF ITR Grant
CCR-0121555, and by the Defense Advanced Research Projects Agency (DARPA).}}
\date{}
\maketitle

\begin{abstract}
The problem of finding a local minimum of a black-box function is central for
understanding local search as well as quantum adiabatic algorithms. \ For
functions on the Boolean hypercube $\left\{  0,1\right\}  ^{n}$, we show a
lower bound of $\Omega\left(  2^{n/4}/n\right)  $ on the number of queries
needed by a quantum computer to solve this problem. \ More surprisingly, our
approach, based on Ambainis's quantum adversary method, also yields a lower
bound of $\Omega\left(  2^{n/2}/n^{2}\right)  $ on the problem's
\textit{classical }randomized query complexity. \ This improves and simplifies
a 1983 result of Aldous. \ Finally, in both the randomized and quantum cases,
we give the first nontrivial lower bounds for finding local minima on grids of
constant dimension $d\geq3$.

\end{abstract}

\section{Introduction\label{INTRO}}

This paper deals with the following problem.

\textsc{Local Search.}\hfill\textit{Given an undirected graph }$G=\left(
V,E\right)  $\textit{\ and a function }$f:V\rightarrow\mathbb{N}$\textit{,
find a local minimum of }$\mathit{f}$\textit{---that is, a vertex }%
$v$\textit{\ such that }$f\left(  v\right)  \leq f\left(  w\right)
$\textit{\ for all neighbors }$w$\textit{ of }$v$\textit{.}

We are interested in the number of \textit{queries} that an algorithm needs to
solve this problem, where a query just returns $f\left(  v\right)  $\ given
$v$. \ We consider deterministic, randomized, and quantum algorithms.
\ Section \ref{MOTIVATION}\ motivates the problem theoretically and
practically; this section explains our results.

We start with some simple observations. \ If $G$ is the complete graph of size
$N$, then clearly $\Omega\left(  N\right)  $\ queries are needed to find a
local minimum (or $\Omega\left(  \sqrt{N}\right)  $\ with a quantum computer
\cite{bbbv}). \ At the other extreme, if $G$\ is a line of length $N$, then
even a deterministic algorithm can find a local minimum in $O\left(  \log
N\right)  $\ queries, using binary search: query the middle two vertices, $v$
and $w$. \ If $f\left(  v\right)  \leq f\left(  w\right)  $, then search the
line of length $\left(  N-2\right)  /2$\ connected to $v$; otherwise search
the line connected to $w$. \ Continue recursively in this manner until a local
minimum is found.

So the interesting case is when $G$ is a graph of `intermediate'
connectedness: for example, the Boolean hypercube $\left\{  0,1\right\}  ^{n}%
$, with two vertices adjacent if and only if they have Hamming distance
$1$.\ \ For this graph, Llewellyn, Tovey, and Trick \cite{ltt}\ showed a
$\Omega\left(  2^{n}/\sqrt{n}\right)  $ lower bound\ on the number of queries
needed by any deterministic algorithm, using a simple adversary argument.
\ Intuitively, until the set of vertices queried so far comprises a
\textit{vertex cut} (that is, splits the graph into two or more connected
components), an adversary is free to return a descending sequence of
$f$-values: $f\left(  v_{1}\right)  =2^{n}$\ for the first vertex $v_{1}%
$\ queried by the algorithm, $f\left(  v_{2}\right)  =2^{n}-1$\ for the second
vertex queried, and so on. \ Moreover, once the set of queried vertices does
comprise a cut, the adversary can choose the largest connected component of
unqueried vertices, and restrict the problem recursively\ to that component.
\ So to lower-bound the deterministic query complexity, it suffices to
lower-bound the size of any cut that splits the graph into two reasonably
large components.\footnote{Llewellyn et al. actually give a tight
characterization of deterministic query complexity in terms of vertex cuts.}
\ For the Boolean hypercube, Llewellyn et al. showed that the best one can do
is essentially to query all $\Omega\left(  2^{n}/\sqrt{n}\right)  $\ vertices
of Hamming weight $n/2$.

Llewellyn et al.'s argument fails completely in the case of randomized
algorithms. \ By Yao's minimax principle, what we want here is a fixed
\textit{distribution} $\mathcal{D}$ over functions $f:\left\{  0,1\right\}
^{n}\rightarrow\mathbb{N}$, such that any deterministic algorithm needs many
queries to find a local minimum of $f$, with high probability if $f$ is drawn
from $\mathcal{D}$. \ Taking $\mathcal{D}$\ to be uniform will not do, since a
local minimum of a uniform random function is easily found. \ However, Aldous
\cite{aldous}\ had the idea of defining $\mathcal{D}$\ via a \textit{random
walk}, as follows. \ Choose a vertex $v_{0}\in\left\{  0,1\right\}  ^{n}%
$\ uniformly at random; then perform an unbiased walk\footnote{Actually,
Aldous used a continuous-time random walk, so the functions would be from
$\left\{  0,1\right\}  ^{n}$\ to $\mathbb{R}$.} $v_{0},v_{1},v_{2},\ldots
$\ starting from $v_{0}$. \ For each vertex $v$, set $f\left(  v\right)  $
equal to the first hitting time of the walk at $v$---that is, $f\left(
v\right)  =\min\left\{  t:v_{t}=v\right\}  $. \ Clearly any $f$ produced in
this way has a unique local minimum at $v_{0}$, since for all $t>0$, if vertex
$v_{t}$\ is visited for the first time at step $t$ then $f\left(
v_{t}\right)  >f\left(  v_{t-1}\right)  $. \ Using sophisticated random walk
analysis, Aldous managed to show a lower bound of $2^{n/2-o\left(  n\right)
}$\ on the expected number of queries needed by any randomized algorithm to
find $v_{0}$.\footnote{Independently and much later, Droste et al.
\cite{djw}\ showed the weaker bound $2^{g\left(  n\right)  }$\ for any
$g\left(  n\right)  =o\left(  n\right)  $.} \ (As we will see in Section
\ref{PRELIM}, this lower bound is close to tight.) \ Intuitively, since a
random walk on the hypercube mixes in $O\left(  n\log n\right)  $\ steps, an
algorithm that has not queried a $v$ with $f\left(  v\right)  <2^{n/2}$\ has
almost no useful information about where the unique minimum $v_{0}$\ is, so
its next query will just be a ``stab in the dark.''

Aldous's result leaves several questions about \textsc{Local Search}%
\ unanswered. \ What if the graph $G$ is a $3$-D cube, on which a random walk
does \textit{not} mix very rapidly? \ Can we still lower-bound the randomized
query complexity of finding a local minimum? \ More generally, what parameters
of $G$ make the problem hard or easy? \ Also, what is the quantum query
complexity of \textsc{Local Search}?

This paper presents a new approach to \textsc{Local Search},\ which we believe
points the way to a complete understanding of its complexity. \ Our approach
is based on the \textit{quantum adversary method}, introduced by Ambainis
\cite{ambainis}\ to prove lower bounds on quantum query complexity.
\ Surprisingly, our approach yields new and simpler lower bounds for the
problem's \textit{classical} randomized query complexity, in addition to
quantum lower bounds. \ Thus, along with recent work by Kerenidis and de Wolf
\cite{kw}\ and by Aharonov and Regev \cite{ar}, this paper illustrates how
quantum ideas can help to resolve classical open problems.

Our results are as follows. \ For the Boolean hypercube $G=\left\{
0,1\right\}  ^{n}$, we show that any quantum algorithm needs $\Omega\left(
2^{n/4}/n\right)  $\ queries to find a local minimum on $G$, and any
randomized algorithm needs $\Omega\left(  2^{n/2}/n^{2}\right)  $\ queries
(improving the $2^{n/2-o\left(  n\right)  }$\ lower bound of Aldous
\cite{aldous}).\ \ Our proofs are elementary and do not require random walk
analysis. \ By comparison, the best known upper bounds are $O\left(
2^{n/3}n^{1/6}\right)  $\ for a quantum algorithm and $O\left(  2^{n/2}%
\sqrt{n}\right)  $\ for a randomized algorithm. \ If $G$ is a $d$-dimensional
grid of size $N^{1/d}\times\cdots\times N^{1/d}$, where $d\geq3$\ is a
constant, then we show that any quantum algorithm needs $\Omega\left(
\sqrt{N^{1/2-1/d}/\log N}\right)  $\ queries to find a local minimum on $G$,
and any randomized algorithm needs $\Omega\left(  N^{1/2-1/d}/\log N\right)
$\ queries. \ No nontrivial lower bounds (randomized or quantum) were
previously known in this case.\footnote{A lower bound on deterministic query
complexity is known for such graphs \cite{lt}.}

In an earlier version of this paper, we raised as our ``most
ambitious''\ conjecture that the deterministic and quantum query complexities
of local search are polynomially related for \textit{every} family of graphs.
\ At the time, it was not even known whether deterministic and
\textit{randomized} query complexities were polynomially related, not even for
simple examples such as the $2$-dimensional square grid. \ Recently Santha and
Szegedy \cite{ss} spectacularly resolved our conjecture, by showing that the
quantum query complexity is at least the $19^{th}$\ root (!) of the
deterministic complexity. \ Given that their result generalizes ours to such
an extent, we feel obligated to defend why this paper is still relevant.
\ First, for specific graphs such as the hypercube, our lower bounds are close
to tight; those of Santha and Szegedy are not. \ Second, we give randomized
lower bounds that are quadratically better than our quantum lower bounds;
Santha and Szegedy give only quantum lower bounds.

In another recent development, Ambainis (personal communication) has improved
our $\Omega\left(  2^{n/4}/n\right)  $\ quantum lower bound for local search
on the hypercube to $2^{n/3}/n^{O\left(  1\right)  }$, using a hybrid
argument. \ Note that Ambainis' lower bound matches the upper bound up to a
polynomial factor.

The paper is organized as follows. \ Section \ref{MOTIVATION}\ motivates lower
bounds on \textsc{Local Search}, pointing out connections to simulated
annealing, quantum adiabatic algorithms, and the complexity class
$\mathsf{TFNP}$\ of total function problems. \ Section \ref{PRELIM}\ defines
notation and reviews basic facts about \textsc{Local Search}, including upper
bounds. \ In Section \ref{ADVERSARY}\ we give an intuitive explanation of
Ambainis's quantum adversary method, then state and prove a classical analogue
of Ambainis's main lower bound theorem. \ Section \ref{SNAKE}\ introduces
\textit{snakes}, a construction by which we apply the two adversary methods to
\textsc{Local Search}.\ \ We show there that to prove lower bounds for any
graph $G$, it suffices to upper-bound a combinatorial parameter $\varepsilon
$\ of a `snake distribution'\ on $G$. \ Section \ref{GRAPHS} applies this
framework to specific examples of graphs: the Boolean hypercube in Section
\ref{BOOLEAN}, and the $d$-dimensional grid in Section \ref{DDIM}.

\section{Motivation\label{MOTIVATION}}

Local search is the most effective weapon ever devised against hard
optimization problems. \ For many real applications, neither backtrack search,
nor approximation algorithms, nor even Grover's algorithm (assuming we had a
quantum computer) can compare. \ Furthermore, along with quantum computing,
local search (broadly defined) is one of the most interesting links between
computer science and Nature. \ It is related to evolutionary biology via
genetic algorithms, and to the physics of materials via simulated annealing.
\ Thus it is both practically and scientifically important to understand its performance.

The conventional wisdom is that, although local search performs well in
practice, its central (indeed defining) flaw is a tendency to get stuck at
local optima. \ If this were correct, one corollary would be that the reason
local search performs so well is that the problem it really solves---finding a
local optimum---is intrinsically easy. \ It would thus be unnecessary to seek
further explanations for its performance. \ Another corollary would be that,
for \textit{unimodal} functions (which have no local optima besides the global
optimum), the global optimum would be easily found.

However, the conventional wisdom is false. \ The results of Llewellyn et al.
\cite{ltt} and Aldous \cite{aldous}\ show that even if $f$ is unimodal, any
classical algorithm that treats $f$ as a black box needs exponential time to
find the global minimum of $f$ in general.\ \ Our results extend this
conclusion to quantum algorithms. \ In our view, the practical upshot of these
results is that they force us to confront the question: What is it about
`real-world'\ problems that makes it easy to find a local optimum? \ That is,
why do exponentially long chains of descending values, such as those used for
lower bounds, almost never occur in practice (even in functions with large
range sizes)?\ \ We do not know a good answer to this.

Our results are also relevant for physics. \ Many physical systems, including
folding proteins and networks of springs and pulleys, can be understood as
performing `local search' through an energy landscape to reach a
locally-minimal energy configuration. \ A key question is, how long will the
system take to reach its ground state (that is, a globally-minimal
configuration)? \ Of course, if there are local optima, the system might
\textit{never} reach its ground state, just as a rock in a mountain crevice
does not roll to the bottom by going up first. \ But what if the energy
landscape is unimodal? \ And moreover, what if the physical system is quantum?
\ Our results show that, for certain energy landscapes, even a quantum system
would take exponential time to reach its ground state, regardless of what
Hamiltonian is applied to it. \ So in particular, the quantum adiabatic
algorithm proposed by Farhi et al. \cite{fggllp}, which can be seen as a
quantum analogue of simulated annealing, needs exponential time to find a
local minimum in the worst case.

Finally, our results have implications for so-called \textit{total function
problems} in complexity theory. \ Megiddo and Papadimitriou \cite{mp}\ defined
a complexity class\footnote{See www.cs.berkeley.edu/\symbol{126}%
aaronson/zoo.html for details about the complexity classes mentioned in this
paper.} $\mathsf{TFNP}$, consisting (informally) of those $\mathsf{NP}%
$\ search problems for which a solution always exists. \ For example, we might
be given a function $f:\left\{  0,1\right\}  ^{n}\rightarrow\left\{
0,1\right\}  ^{n-1}$\ as a Boolean circuit, and asked to find any distinct
$x,y$ pair such that $f\left(  x\right)  =f\left(  y\right)  $. \ This
particular problem belongs to a subclass of $\mathsf{TFNP}$ called
$\mathsf{PPP}$ (Polynomial Pigeonhole Principle). \ Notice that no promise is
involved: the combinatorial nature of the problem itself forces a solution to
exist, even if we have no idea how to find it. \ In a recent talk,
Papadimitriou \cite{papa2} asked broadly whether such `nonconstructive
existence problems' might be good candidates for efficient quantum algorithms.
\ \ In the case of $\mathsf{PPP}$\ problems, the collision lower bound of
Aaronson \cite{aaronson} (improved by Shi \cite{shi}\ and others)\ implies a
negative answer in the black-box setting. \ For other subclasses of
$\mathsf{TFNP}$, such as $\mathsf{PODN}$\ (Polynomial Odd-Degree Node), a
quantum black-box lower bound follows easily from the optimality of Grover's
search algorithm.

However, there is one important subclass of $\mathsf{TFNP}$\ for which no
quantum lower bound was previously known. \ This is $\mathsf{PLS}$ (Polynomial
Local Search), defined by Johnson, Papadimitriou, and Yannakakis \cite{jpy} as
a class of optimization problems whose cost function $f$ and neighborhood
function $\eta$ (that is, the set of neighbors of a given point) are both
computable in polynomial time. \ Given such a problem, the task is to output
any local minimum of the cost function: that is, a $v$ such that $f\left(
v\right)  \leq f\left(  w\right)  $\ for all $w\in\eta\left(  v\right)
$.\ \ The lower bound of Llewellyn et al. \cite{ltt} yields an oracle $A$
relative to which $\mathsf{FP}^{A}\neq\mathsf{PLS}^{A}$, by a standard
diagonalization argument along the lines of Baker, Gill, and Solovay
\cite{bgs}. \ Likewise, the lower bound of Aldous \cite{aldous}\ yields an
oracle relative to which $\mathsf{PLS}\nsubseteq\mathsf{FBPP}$, where
$\mathsf{FBPP}$\ is simply the function version of $\mathsf{BPP}$. \ Our
results yield the first oracle relative to which $\mathsf{PLS}\nsubseteq
\mathsf{FBQP}$. \ In light of this oracle separation, we raise an admittedly
vague question: is there a nontrivial ``combinatorial''\ subclass of
$\mathsf{TFNP}$\ that we can show \textit{is} contained in $\mathsf{FBQP}$?

\section{Preliminaries\label{PRELIM}}

In the \textsc{Local Search}\ problem, we are given an undirected graph
$G=\left(  V,E\right)  $ with $N=\left\vert V\right\vert $,\ and oracle access
to a function $f:V\rightarrow\mathbb{N}$. \ The goal is to find any
\textit{local minimum} of $f$, defined as a vertex $v\in V$ such that
$f\left(  v\right)  \leq f\left(  w\right)  $\ for all neighbors $w$\ of $v$.
\ Clearly such a local minimum exists. \ We want to find one using as few
queries as possible, where a query returns $f\left(  v\right)  $\ given $v$.
\ Queries can be adaptive; that is, can depend on the outcomes of previous
queries. \ We assume $G$ is known in advance, so that only $f$ needs to be
queried. \ Since we care only about query complexity, not computation time,
there is no difficulty in dealing with an infinite range for $f$---though for
our lower bounds, it will turn out that a range of size $O\left(  \left\vert
V\right\vert \right)  $\ suffices.

Our model of query algorithms is the standard one; see \cite{bw}\ for a
survey. \ Given a graph $G$, the deterministic query complexity of
\textsc{Local Search} on $G$, which we denote $\operatorname*{DLS}\left(
G\right)  $, is $\min_{\Gamma}\max_{f}T\left(  \Gamma,f,G\right)  $\ where the
minimum ranges over all deterministic algorithms $\Gamma$, the maximum ranges
over all $f$, and $T\left(  \Gamma,f,G\right)  $\ is the number of queries
made to $f$ by $\Gamma$\ before it halts and outputs a local minimum of $f$
(or $\infty$\ if $\Gamma$ fails to do so). \ The randomized query complexity
$\operatorname*{RLS}\left(  G\right)  $\ is defined similarly, except that now
the algorithm has access to an infinite random string $R$, and must only
output a local minimum with probability at least $2/3$ over $R$. \ For
simplicity, we assume that the number of queries $T$ is the same for all $R$;
clearly this assumption changes the complexity by at most a constant factor.

In the quantum model, an algorithm's state has the form $\sum_{v,z,s}%
\alpha_{v,z,s}\left|  v,z,s\right\rangle $, where $v$ is the label of a vertex
in $G$, and $z$ and $s$ are strings representing the answer register and
workspace respectively. \ The $\alpha_{v,z,s}$'s\ are complex amplitudes
satisfying $\sum_{v,z,s}\left|  \alpha_{v,z,s}\right|  ^{2}=1$. \ Starting
from an arbitrary (fixed) initial state, the algorithm proceeds by an
alternating sequence of \textit{queries} and \textit{algorithm steps}. \ A
query maps each $\left|  v,z,s\right\rangle $\ to $\left|  v,z\oplus f\left(
v\right)  ,s\right\rangle $, where $\oplus$\ denotes bitwise exclusive-OR.
\ An algorithm step multiplies the vector of $\alpha_{v,z,s}$'s\ by an
arbitrary unitary matrix that does not depend on $f$. \ Letting $\mathcal{M}%
_{f}$\ denote the set of local minima of $f$, the algorithm succeeds if at the
end $\sum_{v,z,s~:~v\in\mathcal{M}_{f}}\left|  \alpha_{v,z,s}\right|  ^{2}%
\geq\frac{2}{3}$. \ Then the bounded-error quantum query complexity, or
$\operatorname*{QLS}\left(  G\right)  $, is defined as the minimum number of
queries used by a quantum algorithm that succeeds on every $f$.

It is immediate that $\operatorname*{QLS}\left(  G\right)  \leq
\operatorname*{RLS}\left(  G\right)  \leq\operatorname*{DLS}\left(  G\right)
\leq N$. \ Also, letting $\delta$\ be the maximum degree of $G$, we have the
following trivial lower bound. \vspace{-0.1in}
\begin{proposition}
\label{degree}$\operatorname*{RLS}\left(  G\right)  =\Omega\left(
\delta\right)  $\ and $\operatorname*{QLS}\left(  G\right)  =\Omega\left(
\sqrt{\delta}\right)  $.
\end{proposition}
\vspace{-0.2in} \begin{proof}%
Let $v$ be a vertex of $G$ with degree $\delta$. \ Choose a neighbor $w$ of
$v$ uniformly at random, and let $f\left(  w\right)  =1$. \ Let $f\left(
v\right)  =2$, and $f\left(  u\right)  =3$ for all neighbors $u$ of $v$ other
than $w$. \ Let $S$\ be the neighbor set of $v$ (including $v$ itself); then
for all $x\notin S$, let $f\left(  x\right)  =3+\Delta\left(  x,S\right)
$\ where $\Delta\left(  x,S\right)  $ is the minimum distance from $x$ to a
vertex in $S$. \ Clearly $f$ has a unique local minimum at $w$. \ However,
finding $y$ requires exhaustive search among the $\delta$\ neighbors of $v$,
which requires $\Omega\left(  \sqrt{\delta}%
\right)  $ quantum queries
\cite{bbbv}.
\end{proof}
 \vspace{-0.1in} A
corollary of Proposition \ref{degree}\ is that classically, zero-error
randomized query complexity is equivalent to bounded-error up to a constant
factor. \ For given a candidate local minimum $v$, one can check using
$O\left(  \delta\right)  $\ queries that $v$ is indeed a local minimum.
\ Since $\Omega\left(  \delta\right)  $\ queries are needed anyway, this
verification step does not affect the overall complexity.

As pointed out by Aldous \cite{aldous}, a classical randomized algorithm can
find a local minimum of $f$ with high probability in $O\left(  \sqrt{N\delta
}\right)  $\ queries. \ The algorithm just queries $\sqrt{N\delta}$\ vertices
uniformly at random, and lets $v_{0}$\ be a queried vertex for which $f\left(
v\right)  $\ is minimal. \ It then follows $v_{0}$\ to a local minimum by
steepest descent. \ That is, for $t=0,1,2,\ldots$, it queries all neighbors of
$v_{t}$, halts if $v_{t}$\ is a local minimum, and otherwise sets $v_{t+1}%
$\ to be the neighbor $w$ of $v_{t}$\ for which $f\left(  w\right)  $\ is
minimal (breaking ties by lexicographic ordering). \ A similar idea yields an
improved quantum upper bound. \vspace{-0.1in}

\begin{proposition}
\label{upper}For any $G$, $\operatorname*{QLS}\left(  G\right)  =O\left(
N^{1/3}\delta^{1/6}\right)  $.
\end{proposition}

\vspace{-0.2in} \begin{proof}
The algorithm first chooses $N^{2/3}%
\delta^{1/3}%
$\ vertices of $G$ uniformly
at random, then uses Grover search to find a chosen vertex $v_{0}%
$ for which
$f\left(  v\right)  $\ is minimal. \ By a result of D\"{u}%
rr and H\o yer
\cite{dh}%
, this can be done with high probability in $O\left(  N^{1/3}%
\delta^{1/6}%
\right)  $\ queries. \ Next, for $t=0,1,2,\ldots$, the algorithm
performs Grover search over all neighbors of $v_{t}%
$, looking for a neighbor
$w$ such that $f\left(  w\right)  <f\left(  v_{t}%
\right)  $. \ If it finds
such a $w$, then it sets $v_{t+1}%
:=w$ and continues to the next iteration.
\ Otherwise, it repeats the Grover search $\log\left(  N/\delta\right)
$\ times before finally giving up and returning $v_{t}%
$\ as a claimed local minimum.

The expected number of vertices $u$ such that $f\left(  u\right)  <f\left(
v_{0}%
\right)  $ is at most $N/\left(  N^{2/3}\delta^{1/3}%
\right)  =\left(
N/\delta\right)  ^{1/3}$. \ Since $f\left(  v_{t+1}%
\right)  <f\left(
v_{t}%
\right)  $ for all $t$, clearly the number of such $u$ provides an upper
bound on $t$. \ Furthermore, assuming there exists a $w$ such that $f\left(
w\right)  <f\left(  v_{t}%
\right)  $, the expected number of repetitions of
Grover's algorithm until such a $w$ is found is $O\left(  1\right)  $. \ Since
each repetition takes $O\left(  \sqrt{\delta}%
\right)  $\ queries, by linearity
of expectation the total expected number of queries used by the algorithm is
therefore%
\[
O\left(  N^{1/3}%
\delta^{1/6}+\left(  N/\delta\right)  ^{1/3}\sqrt{\delta}%
+\log\left(  N/\delta\right)  \sqrt{\delta}%
\right)
\]
or $O\left(  N^{1/3}\delta^{1/6}%
\right)  $. \ To see that the algorithm finds
a local minimum with high probability, observe that for each $t$, the
probability of not finding a $w$ such that $f\left(  w\right)  <f\left(
v_{t}%
\right)  $, given that one exists, is at most $c^{-\log\left(
N/\delta\right)  }%
\leq\left(  \delta/N\right)  ^{1/3}%
/10$\ for a suitable
constant $c$. \ So by the union bound, the probability that the algorithm
returns a `false positive' is at most $\left(  N/\delta\right)^{1/3}%
\cdot\left(  \delta/N\right)  ^{1/3}/10=1/10$.
\end{proof}
 
\section{Relational Adversary Method\label{ADVERSARY}}

We know of essentially two methods for proving lower bounds on quantum query
complexity: the polynomial method of Beals et al. \cite{bbcmw}, and the
quantum adversary method of Ambainis \cite{ambainis}.\footnote{We are thinking
here of the hybrid method \cite{bbbv}\ as a cousin of the adversary method.}
\ For a few problems, such as the collision problem \cite{aaronson}, the
polynomial method succeeded where the adversary method failed. \ However, for
problems that lack permutation symmetry (such as \textsc{Local Search}), the
adversary method has proven more effective.\footnote{Indeed, Ambainis
\cite{ambainis0}\ has given problems for which the adversary method provably
yields a better lower bound than the polynomial method.}

How could a quantum lower bound method possibly be applied classically? \ When
proving randomized lower bounds, the tendency is to attack ``bare-handed'':
fix a distribution over inputs, and let $x_{1},\ldots,x_{t}$\ be the locations
queried so far by the algorithm. \ Show that for small $t$, the posterior
distribution over inputs, \textit{conditioned} on $x_{1},\ldots,x_{t}$, is
still `hard' with high probability---so that the algorithm knows almost
nothing even about which location $x_{t+1}$\ to query next. \ This is
essentially the approach taken by Aldous \cite{aldous}\ to prove a
$2^{n/2-o\left(  n\right)  }$\ lower bound on $\operatorname*{RLS}\left(
\left\{  0,1\right\}  ^{n}\right)  $.

In the quantum case, however, it is unclear how to specify what an algorithm
`knows'\ after a given number of queries. \ So we are almost \textit{forced}
to step back, and identify general combinatorial properties of input sets that
make them hard to distinguish. \ Once we have such properties, we can then try
to exhibit them in functions of interest.

We believe this \textquotedblleft gloved\textquotedblright\ attack can be
useful for classical lower bounds as well as quantum ones. \ In our
\textit{relational adversary method}, we assume there exists a $T$-query
randomized algorithm for function $F$. \ We consider a set $\mathcal{A}$\ of
$0$-inputs of $F$, a set $\mathcal{B}$\ of $1$-inputs, and an arbitrary
real-valued \textit{relation function} $R\left(  A,B\right)  \geq0$\ for
$A\in\mathcal{A}$\ and $B\in\mathcal{B}$. \ Intuitively, $R\left(  A,B\right)
$\ should be large if $A$ and $B$ differ in only a few locations. \ We then
fix a probability distribution $\mathcal{D}$\ over inputs; by Yao's minimax
principle, there exists a $T$-query deterministic algorithm $\Gamma^{\ast}%
$\ that succeeds with high probability on inputs drawn from $\mathcal{D}$.
\ Let $W_{A}$\ be the set of $0$-inputs and $W_{B}$\ the set of $1$-inputs on
which $\Gamma^{\ast}$\ succeeds. \ Using the relation function $R$, we define
a \textit{separation measure} $S$\ between $W_{A}$\ and $W_{B}$, and show that
(1) initially $S=0$,\ (2) by the end of the computation $S$\ must be large,
and (3) $S$ increases by only a small amount as the result of each query. \ It
follows that $T$\ must be large.

Undoubtedly any randomized lower bound proved using our relational method
could also be proved \textquotedblleft bare-handed,\textquotedblright\ without
any quantum intuition. \ However, our method makes it easier to focus on what
is unique about a problem, and ignore what is common among many problems.

Our starting point is the ``most general''\ adversary theorem in Ambainis's
original paper (Theorem 6\ in \cite{ambainis}), which he introduced to prove a
quantum lower bound for the problem of inverting a permutation. \ Here the
input is a permutation $\sigma\left(  1\right)  ,\ldots,\sigma\left(
N\right)  $, and the task is to output $0$ if $\sigma^{-1}\left(  1\right)
\leq N/2$\ and $1$ otherwise. \ To lower-bound this problem's query
complexity, what we would like to say is this:

\textit{Given any }$0$\textit{-input }$\sigma$\textit{\ and any location }%
$x$\textit{, if we choose a random }$1$\textit{-input }$\tau$\textit{\ that is
`related' to }$\sigma$\textit{, then the probability }$\theta\left(
\sigma,x\right)  $\textit{ over }$\tau$\textit{\ that }$\sigma\left(
x\right)  $ \textit{does not equal }$\tau\left(  x\right)  $\textit{\ is
small. \ In other words, the algorithm is unlikely to distinguish }$\sigma
$\textit{\ from a random neighbor }$\tau$\textit{\ of }$\sigma$\textit{\ by
querying }$x$\textit{.}

Unfortunately, the above claim is false. \ Letting $x=\sigma^{-1}\left(
1\right)  $, we have that $\sigma\left(  x\right)  \neq\tau\left(  x\right)
$\ for \textit{every} $1$-input $\tau$, and thus $\theta\left(  \sigma
,x\right)  =1$. \ Ambainis resolves this difficulty by letting us take the
maximum, over all $0$-inputs $\sigma$\ and $1$-inputs $\tau$ that are related
and differ at $x$, of the \textit{geometric mean} $\allowbreak\sqrt
{\theta\left(  \sigma,x\right)  \theta\left(  \tau,x\right)  }$. \ Even if
$\theta\left(  \sigma,x\right)  =1$, the geometric mean is still small
provided that $\theta\left(  \tau,x\right)  $\ is small. \ More formally:

\begin{theorem}
[Ambainis]\label{ambthm}Let $\mathcal{A}\subseteq F^{-1}\left(  0\right)  $
and $\mathcal{B}\subseteq F^{-1}\left(  1\right)  $\ be sets of inputs to
function $F$. \ Let $R\left(  A,B\right)  \geq0$ be a real-valued function,
and for $A\in\mathcal{A}$, $B\in\mathcal{B}$, and location $x$, let%
\begin{align*}
\theta\left(  A,x\right)   &  =\frac{\sum_{B^{\ast}\in\mathcal{B}~:~A\left(
x\right)  \neq B^{\ast}\left(  x\right)  }R\left(  A,B^{\ast}\right)  }%
{\sum_{B^{\ast}\in\mathcal{B}}R\left(  A,B^{\ast}\right)  },\\
\theta\left(  B,x\right)   &  =\frac{\sum_{A^{\ast}\in\mathcal{A}~:~A^{\ast
}\left(  x\right)  \neq B\left(  x\right)  }R\left(  A^{\ast},B\right)  }%
{\sum_{A^{\ast}\in\mathcal{A}}R\left(  A^{\ast},B\right)  },
\end{align*}
where the denominators are all nonzero. \ Then the number of quantum queries
needed to evaluate $F$ with at least $9/10$ probability is $\Omega\left(
1/\upsilon_{\operatorname*{geom}}\right)  $, where%
\[
\upsilon_{\operatorname*{geom}}=\max_{\substack{A\in\mathcal{A},~B\in
\mathcal{B},~x~:\\R\left(  A,B\right)  >0,~A\left(  x\right)  \neq B\left(
x\right)  }}\sqrt{\theta\left(  A,x\right)  \theta\left(  B,x\right)  }.
\]
\end{theorem}
To illustrate we show the following.
\begin{proposition}
[Ambainis]The quantum query complexity of inverting a permutation is
$\Omega\left(  \sqrt{N}\right)  $.
\end{proposition}
\vspace{-0.1in}
\begin{proof}
\vspace{-0.1in}
Let $\mathcal{A}%
$\ be the set of all permutations $\sigma$\ with $\sigma
^{-1}%
\left(  1\right)  $ $\leq N/2$, and $\mathcal{B}%
$\ be the set of
permutations $\tau$\ with $\tau^{-1}%
\left(  1\right)  >N/2$. \ Given
$\sigma\in\mathcal{A}%
$\ and $\tau\in\mathcal{B}%
$, let $R\left(  \sigma
,\tau\right)  =1$\ if $\sigma$\ and $\tau$\ differ only at locations
$\sigma^{-1}%
\left(  1\right)  $\ and $\tau^{-1}%
\left(  1\right)  $, and
$R\left(  \sigma,\tau\right)  =0$\ otherwise. \ Then given $\sigma,\tau$\ with
$R\left(  \sigma,\tau\right)  =1$,\ if $x\neq\sigma^{-1}%
\left(  1\right)
$\ then $\theta\left(  \sigma,x\right)  =2/N$, and if $x\neq\tau^{-1}%
\left(
1\right)  $\ then $\theta\left(  \tau,x\right)  =2/N$. \ So $\max
_{x~:~\sigma\left(  x\right)  \neq\tau\left(  x\right)  }%
\sqrt{\theta\left(
\sigma,x\right)  \theta\left(  \tau,x\right)  }%
=\sqrt{2/N}$.
\end{proof}
 The only difference between Theorem
\ref{ambthm}\ and our relational adversary theorem is that in the latter, we
take the \textit{minimum} of $\theta\left(  A,x\right)  $ and $\theta\left(
B,x\right)  $\ instead of the geometric mean. \ Taking the reciprocal then
gives up to a quadratically better lower bound: for example, we obtain that
the randomized query complexity of inverting a permutation is $\Omega\left(
N\right)  $. \ However, the proofs of the two theorems are quite different.
\begin{theorem}
\label{classadv}Let $\mathcal{A},\mathcal{B},R,\theta$ be as in Theorem
\ref{ambthm}. \ Then the number of randomized queries needed to evaluate
$F$\ with at least $9/10$ probability is $\Omega\left(  1/\upsilon_{\min
}\right)  $, where%
\[
\upsilon_{\min}=\max_{\substack{A\in\mathcal{A},~B\in\mathcal{B}%
,~x~:\smallskip\,\\R\left(  A,B\right)  >0,~A\left(  x\right)  \neq B\left(
x\right)  }}\min\left\{  \theta\left(  A,x\right)  ,\theta\left(  B,x\right)
\right\}  .
\]
\end{theorem}
\vspace{-0.5in}
\begin{proof}%
Let $\Gamma$ be a randomized algorithm that, given an input $A$, returns
$F\left(  A\right)  $ with at least $9/10$ probability. \ Let $T$ be the
number of queries made by $\Gamma$. \ For all $A\in\mathcal{A}%
$,
$B\in\mathcal{B}$, define%
\begin{align*}%
M\left(  A\right)   &  =\sum_{B^{\ast}\in\mathcal{B}}R\left(  A,B^{\ast
}%
\right)  ,\\
M\left(  B\right)   &  =\sum_{A^{\ast}\in\mathcal{A}%
}R\left(  A^{\ast
},B\right)  ,\\
M  &  =\sum_{A^{\ast}\in\mathcal{A}%
}M\left(  A^{\ast}\right)  =\sum_{B^{\ast
}\in\mathcal{B}}M\left(  B^{\ast}%
\right)  .
\end{align*}
Now let $\mathcal{D}_{A}%
$\ be the distribution over $A\in\mathcal{A}%
$\ in
which each $A$ is chosen with probability $M\left(  A\right)  /M$; and let
$\mathcal{D}%
_{B}$\ be the distribution over $B\in\mathcal{B}%
$\ in which each
$B$ is chosen with probability $M\left(  B\right)  /M$. \ Let $\mathcal{D}%
$\ be an equal mixture of $\mathcal{D}_{A}$\ and $\mathcal{D}%
_{B}%
$. \ By Yao's
minimax principle, there exists a deterministic algorithm $\Gamma^{\ast}%
$\ that makes $T$ queries, and succeeds with at least $9/10$ probability given
an input drawn from $\mathcal{D}%
$. \ Therefore $\Gamma^{\ast}%
$\ succeeds with
at least $4/5$\ probability given an input drawn from $\mathcal{D}%
_{A}$ alone,
or from $\mathcal{D}_{B}%
$ alone. \ In other words, letting $W_{A}$\ be the set
of $A\in\mathcal{A}%
$\ and $W_{B}$\ the set of $B\in\mathcal{B}$\ on which
$\Gamma^{\ast}%
$\ succeeds, we have%
\[
\sum_{A\in W_{A}}M\left(  A\right)  \geq\frac{4}%
{5}M,\,\,\,\,\,\,\sum_{B\in
W_{B}}M\left(  B\right)  \geq\frac{4}%
{5}M.
\]
Define a predicate $P^{\left(  t\right)  }%
\left(  A,B\right)  $, which is true
if $\Gamma^{\ast}%
$ has distinguished $A\in\mathcal{A}$ from $B\in\mathcal{B}%
$\ by the $t^{th}%
$\ query and false otherwise. \ (To distinguish $A$ from $B$
means to query an index $x$\ for which $A\left(  x\right)  \neq B\left(
x\right)  $, given either $A$ or $B$ as input.) \ Also, for all $A\in
\mathcal{A}%
$, define a score function%
\[
S^{\left(  t\right)  }%
\left(  A\right)  =\sum_{B^{\ast}\in\mathcal{B}%
~:~P^{\left(  t\right)  }%
\left(  A,B^{\ast}\right)  }R\left(  A,B^{\ast
}%
\right).
\]
This function measures how much ``progress'' has been made so far in
separating $A$ from $\mathcal{B}%
$-inputs, where the $\mathcal{B}%
$-inputs\ are
weighted by $R\left(  A,B\right)  $. \ Similarly, for all $B\in\mathcal{B}%
$
define%
\[
S^{\left(  t\right)  }\left(  B\right)  =\sum_{A^{\ast}%
\in\mathcal{A}%
~:~P^{\left(  t\right)  }\left(  A^{\ast},B\right)  }%
R\left(  A^{\ast
}%
,B\right).
\]
It is clear that for all $t$,%
\[
\sum_{A\in\mathcal{A}%
}S^{\left(  t\right)  }\left(  A\right)  =\sum
_{B\in\mathcal{B}%
}S^{\left(  t\right)  }%
\left(  B\right)  .
\]
So we can denote the above sum by $S^{\left(  t\right)  }%
$ and think of it as
a\ global progress measure. \ We will show the following about $S^{\left(
t\right)  }%
$:
\vspace{-0.1in}
\begin{enumerate}
\item[(i)] $S^{\left(  0\right)  }%
=0$ initially.
\item[(ii)] $S^{\left(  T\right)  }%
\geq3M/5$ by the end.
\item[(iii)] $\Delta S^{\left(  t\right)  }%
\leq3\upsilon_{\min}M$\ for all
$t$, where $\Delta S^{\left(  t\right)  }%
=S^{\left(  t\right)  }-S^{\left(
t-1\right)  }%
$ is the amount by which $S^{\left(  t\right)  }%
$\ increases as
the result of a single query.
\end{enumerate}%
\vspace{-0.1in}
It follows from (i)-(iii) that%
\[
T\geq\frac{3M/5}%
{3\upsilon_{\min}M}=\frac{1}{5\upsilon_{\min}}%
\]
which establishes the theorem. \ Part (i) is obvious. \ For part (ii), observe
that for every pair $\left(  A,B\right)  $ with $A\in W_{A}%
$ and $B\in W_{B}$,
the algorithm $\Gamma^{\ast}%
$\ must query an $x$ such that $A\left(  x\right)
\neq B\left(  x\right)  $. \ Thus%
\begin{align*}%
S^{\left(  T\right)  }  &  \geq\sum_{A\in W_{A},~B\in W_{B}}%
R\left(
A,B\right) \\
&  \geq\sum_{A\in W_{A}}%
M\left(  A\right)  -\sum_{B\notin W_{B}}M\left(
B\right)  \geq\frac{4}%
{5}M-\frac{1}{5}M.
\end{align*}%
It remains only to show part (iii). \ Suppose $\Delta S^{\left(  t\right)
}%
>3\upsilon_{\min}%
M$\ for some $t$; we will obtain a contradiction.\ \ Let%
\[
\Delta S^{\left(  t\right)  }%
\left(  A\right)  =S^{\left(  t\right)  }%
\left(
A\right)  -S^{\left(  t-1\right)  }%
\left(  A\right)  ,
\]
and let $C_{A}$\ be the set of $A\in\mathcal{A}%
$ for which $\Delta S^{\left(
t\right)  }\left(  A\right)  >\upsilon_{\min}%
M\left(  A\right)  $. \ Since%
\[
\sum_{A\in\mathcal{A}}%
\Delta S^{\left(  t\right)  }\left(  A\right)  =\Delta
S^{\left(  t\right)  }%
>3\upsilon_{\min}%
M,
\]
it follows by Markov's inequality that%
\[
\sum_{A\in C_{A}%
}\Delta S^{\left(  t\right)  }\left(  A\right)  \geq
\frac{2}{3}%
\Delta S^{\left(  t\right)  }.
\]
Similarly, letting $C_{B}%
$\ be the set of $B\in\mathcal{B}$ for which $\Delta
S^{\left(  t\right)  }%
\left(  B\right)  >\upsilon_{\min}%
M\left(  B\right)  $,
we have%
\[
\sum_{B\in C_{B}}%
\Delta S^{\left(  t\right)  }\left(  B\right)  \geq
\frac{2}{3}%
\Delta S^{\left(  t\right)  }.
\]
In other words, at least $2/3$\ of the increase in $S^{\left(  t\right)  }%
$\ comes from $\left(  A,B\right)  $\ pairs such that $A\in C_{A}%
$,\ and at
least $2/3$\ comes from $\left(  A,B\right)  $\ pairs such that $B\in C_{B}%
$.
\ Hence, by a `pigeonhole' argument, there exists an $A\in C_{A}%
$ and $B\in
C_{B}%
$\ with $R\left(  A,B\right)  >0$\ that are distinguished by the $t^{th}%
$\ query. \ In other words, there exists an $x$ with $A\left(  x\right)  \neq
B\left(  x\right)  $, such that the $t^{th}%
$\ index queried by $\Gamma^{\ast}%
$\ is $x$ whether the input is $A$ or $B$. \ Then since $A\in C_{A}%
$, we have
$\upsilon_{\min}M\left(  A\right)  <\Delta S^{\left(  t\right)  }%
\left(
A\right)  $, and hence%
\[
\upsilon_{\min}%
<\frac{\Delta S^{\left(  t\right)  }\left(  A\right)
}{M\left(  A\right)  }%
\leq\frac{\sum_{B^{\ast}\in\mathcal{B}~:~A\left(
x\right)  \neq B^{\ast}%
\left(  x\right)  }R\left(  A,B^{\ast}\right)  }%
{\sum_{B^{\ast}%
\in\mathcal{B}}R\left(  A,B^{\ast}\right)  }%
\]
which equals $\theta\left(  A,x\right)  $. \ Similarly $\upsilon_{\min}%
<\theta\left(  B,x\right)  $ since $B\in C_{B}%
$. \ This contradicts the
definition%
\[
\upsilon_{\min}%
=\max_{\substack{A\in\mathcal{A},~B\in\mathcal{B}%
,~x~:\smallskip\,\\R\left(  A,B\right)  >0,~A\left(  x\right)  \neq B\left(
x\right)  }%
}%
\min\left\{  \theta\left(  A,x\right)  ,\theta\left(  B,x\right)
\right\}  ,
\]
and we are done.
\end{proof}%

\section{Snakes\label{SNAKE}}

For our lower bounds, it will be convenient to generalize random walks to
arbitrary distributions over paths, which we call \textit{snakes}. \vspace{-0.1in}

\begin{definition}
\label{snake}Given a vertex $h$ in $G$ and a positive integer $L$, a
\textit{snake distribution} $\mathcal{D}_{h,L}$\ (parameterized by $h$ and
$L$) is a probability distribution over paths $\left(  x_{0},\ldots
,x_{L-1}\right)  $\ in $G$, such that each $x_{t}$\ is either equal or
adjacent to $x_{t+1}$, and $x_{L-1}=h$. \ Let $D_{h,L}$\ be the support of
$\mathcal{D}_{h,L}$. \ Then an element of $D_{h,L}$\ is called a
\textit{snake}; the part near $x_{0}$\ is the \textit{tail} and the part near
$x_{L-1}=h$\ is the \textit{head}.
\end{definition}

\vspace{-0.1in} Given a snake $X$ and integer $t$, we use $X\left[  t\right]
$\ as shorthand for $\left\{  x_{0},\ldots,x_{t}\right\}  $.

\begin{definition}
\label{elgood}We say a snake $X\in D_{h,L}$\ is $\varepsilon$\textit{-good}%
\ if the following holds. \ Choose $j$ uniformly at random from $\left\{
0,\ldots,L-1\right\}  $, and let $Y=\left(  y_{0},\ldots,y_{L-1}\right)  $\ be
a snake drawn from $\mathcal{D}_{h,L}$\ conditioned on $x_{t}=y_{t}$\ for all
$t>j$. \ Then
\begin{enumerate}
\item[(i)] Letting $S_{X,Y}$\ be the set of vertices $v$ in $X\cap Y$\ such
that $\min\left\{  t:x_{t}=v\right\}  =\min\left\{  t:y_{t}=v\right\}  $, we
have%
\[
\Pr_{j,Y}\left[  X\cap Y=S_{X,Y}\right]  \geq9/10.
\]
\item[(ii)] For all vertices $v$, $\Pr_{j,Y}\left[  v\in Y\left[  j\right]
\right]  \leq\varepsilon$.
\end{enumerate}
\end{definition}

The procedure above---wherein we choose a $j$ uniformly at random, then draw a
$Y$\ from $\mathcal{D}_{h,L}$\ consistent with $X$ on all steps later than
$j$---will be important in what follows.\ \ We call it \textit{the snake
}$\mathit{X}$\textit{ flicking its tail}. \ Intuitively, a snake is good if it
is spread out fairly evenly in $G$---so that when it flicks its tail, (1) with
high probability the old and new tails do not intersect, and (2) any
particular vertex is hit by the new tail with probability at most
$\varepsilon$.

We now explain our `snake method' for proving lower bounds for \textsc{Local
Search}. \ Given a snake $X$, we define an input $f_{X}$\ with a unique local
minimum at $x_{0}$, and $f$-values that decrease along $X$ from head to tail.
\ Then, given inputs $f_{X}$ and $f_{Y}$ with $X\cap Y=S_{X,Y}$, we let the
relation function $R\left(  f_{X},f_{Y}\right)  $ be proportional to the
probability that snake $Y$ is obtained by $X$ flicking its tail. \ (If $X\cap
Y\neq S_{X,Y}$ we let $R=0$.) \ Let $f_{X}$ and $g_{Y}$ be inputs with
$R\left(  f_{X},g_{Y}\right)  >0$, and let $v$ be a vertex such that
$f_{X}\left(  v\right)  \neq g_{Y}\left(  v\right)  $. \ Then if all snakes
were good, there would be two mutually exclusive cases: (1) $v$ belongs to the
tail of $X$, or (2) $v$ belongs to the tail of $Y$. \ In case (1), $v$ is hit
with small probability when $Y$\ flicks its tail, so $\theta\left(
f_{Y},v\right)  $\ is small. \ In case (2), $v$ is hit with small probability
when $X$ flicks its tail, so $\theta\left(  f_{X},v\right)  $\ is small. \ In
either case, then, the \textit{geometric mean} $\sqrt{\theta\left(
f_{X},v\right)  \theta\left(  f_{Y},v\right)  }$\ and \textit{minimum}
$\min\left\{  \theta\left(  f_{X},v\right)  ,\theta\left(  f_{Y},v\right)
\right\}  $\ are small. \ So even though $\theta\left(  f_{X},v\right)  $\ or
$\theta\left(  f_{Y},v\right)  $\ could be large individually, Theorems
\ref{ambthm}\ and \ref{classadv} yield a good lower bound, as in the case of
inverting a permutation (see Figure 1).%
\begin{figure}
[ptb]
\begin{center}
\includegraphics[
trim=3.754344in 5.368080in 3.618745in 0.393184in,
height=2.3168in,
width=3.4532in
]%
{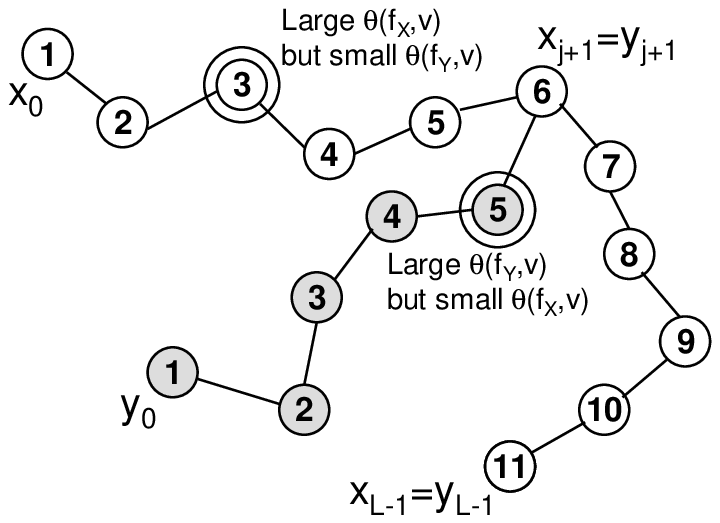}%
\caption{For every vertex $v$ such that $f_{X}\left(  v\right)  \neq
f_{Y}\left(  v\right)  $, either when snake $X$ flicks its tail $v$ is not hit
with high probability, or when snake $Y$ flicks its tail $v$ is not hit with
high probability.}%
\label{snakefig}%
\end{center}
\end{figure}

One difficulty is that not all snakes are good; at best, a large fraction of
them are. \ We could try deleting all inputs $f_{X}$\ such that $X$ is not
good, but that might ruin some remaining inputs, which would then have fewer
neighbors. \ So we would have to delete \textit{those} inputs as well, and so
on ad infinitum. \ What we need is basically a way to replace ``all
inputs''\ by ``most inputs''\ in Theorems \ref{ambthm}\ and \ref{classadv}.

Fortunately, a simple graph-theoretic lemma can accomplish this. \ The lemma
(see Diestel \cite[p.6]{diestel} for example) says that any graph with average
degree at least $k$ contains an induced subgraph with \textit{minimum} degree
at least $k/2$. \ Here we prove a weighted analogue of the lemma.
\vspace{-0.1in}
\begin{lemma}
\label{subgraph}Let $p\left(  1\right)  ,\ldots,p\left(
m\right)  $\ be positive reals summing to $1$. \ Also let $w\left(
i,j\right)  $ for $i,j\in\left\{  1,\ldots,m\right\}  $\ be nonnegative reals
satisfying $w\left(  i,j\right)  =w\left(  j,i\right)  $\ and $\allowbreak
\sum_{i,j}w\left(  i,j\right)  \geq r$. \ Then there exists a nonempty subset
$U\subseteq\left\{  1,\ldots,m\right\}  $\ such that for all $i\in U$,
$\sum_{j\in U}w\left(  i,j\right)  \geq rp\left(  i\right)  /2.$
\end{lemma}
\vspace{-0.1in}
\begin{proof}%
If $r=0$ then the lemma trivially holds, so assume $r>0$.\ \ We construct $U$
via an iterative procedure. \ Let $U\left(  0\right)  =\left\{  1,\ldots
,m\right\}  $. \ Then for all $t$, if there exists an\ $i^{\ast}%
\in U\left(
t\right)  $ for which%
\[
\sum_{j\in U\left(  t\right)  }%
w\left(  i^{\ast},j\right)  <\frac{r}%
{2}p\left(  i^{\ast}%
\right)  ,
\]
then set $U\left(  t+1\right)  =U\left(  t\right)  \setminus\left\{  i^{\ast
}%
\right\}  $. \ Otherwise halt and return $U=U\left(  t\right)  $. \ To see
that the $U$ so constructed is nonempty, observe that when we remove $i^{\ast
}%
$, the sum $\sum_{i\in U\left(  t\right)  }%
p\left(  i\right)  $\ decreases by
$p\left(  i^{\ast}%
\right)  $, while $\sum_{i,j\in U\left(  t\right)  }%
w\left(
i,j\right)  $\ decreases by at most%
\[
\sum_{j\in U\left(  t\right)  }%
w\left(  i^{\ast},j\right)  +\sum_{j\in
U\left(  t\right)  }%
w\left(  j,i^{\ast}\right)  <rp\left(  i^{\ast}%
\right)  .
\]
So since $\sum_{i,j\in U\left(  t\right)  }%
w\left(  i,j\right)  $\ was
positive to begin with, it must still be positive at the end of the procedure;
hence $U$\ must be nonempty.
\end{proof}%
\vspace{-0.1in}
We can now prove the main result of the section.
\vspace{-0.1in}
\begin{theorem}
\vspace{-0.1in}
\label{kappathm}Suppose a snake drawn from $\mathcal{D}_{h,L}$\ is
$\varepsilon$\textit{-}good\ with probability at least $9/10$. \ Then%
\[
\operatorname*{RLS}\left(  G\right)  =\Omega\left(  1/\varepsilon\right)
,~~~~~~\operatorname*{QLS}\left(  G\right)  =\Omega\left(  \sqrt
{1/\varepsilon}\right)  .
\]
\end{theorem}
\begin{proof}
Given a snake $X\in D_{h,L}%
$, we construct an input function $f_{X}%
$ as
follows.\ \ For each $v\in X$, let $f_{X}%
\left(  v\right)  =\min\left\{
t:x_{t}%
=v\right\}  $; and for each $v\notin X$, let $f_{X}%
\left(  v\right)
=\Delta\left(  v,h\right)  +L$\ where $\Delta\left(  v,h\right)  $\ is the
distance from $v$ to $h$ in $G$. \ Clearly $f_{X}%
$ so defined has a unique
local minimum at $x_{0}%
$. \ To obtain a decision problem, we stipulate that
querying $x_{0}%
$\ reveals an answer bit ($0$ or $1$) in addition to
$f_{X}\left(  x_{1}%
\right)  $; the algorithm's goal is then to return the
answer bit. \ Obviously a lower bound for the decision problem implies a
corresponding lower bound for the search problem. Let us first prove the
theorem in the case that all snakes in $D_{h,L}%
$\ are $\varepsilon$%
\textit{-}%
good. \ Let $p\left(  X\right)  $\ be the probability of drawing
snake $X$ from $\mathcal{D}%
_{h,L}%
$. \ Also, given snakes $X,Y$\ and
$j\in\left\{  0,\ldots,L-1\right\}  $, let $q_{j}%
\left(  X,Y\right)  $\ be the
probability that $X^{\ast}=Y$, if $X^{\ast}%
$ is drawn from $\mathcal{D}_{h,L}%
$
conditioned on agreeing with $X$\ on all steps later than $j$. \ Then define%
\[
w\left(  X,Y\right)  =\frac{p\left(  X\right)  }%
{L}\sum_{j=0}^{L-1}%
q_{j}%
\left(  X,Y\right)  .
\]
Our first claim is that $w$\ is symmetric; that is, $w\left(  X,Y\right)
=w\left(  Y,X\right)  $. \ It suffices to show that
\[
p\left(  X\right)  q_{j}%
\left(  X,Y\right)  =p\left(  Y\right)  q_{j}%
\left(
Y,X\right)
\]
for all $j$. \ We can assume $X$\ agrees with $Y$ on all steps later than $j$,
since otherwise $q_{j}%
\left(  X,Y\right)  =q_{j}\left(  Y,X\right)  =0$.
\ Given an $X^{\ast}%
\in D_{h,L}$, let $A$ denote the event that $X^{\ast}%
$
agrees with $X$ (or equivalently $Y$) on all steps later than $j$, and let
$B_{X}%
$ (resp. $B_{Y}$) denote the event that $X^{\ast}%
$ agrees with
$X$\ (resp. $Y$) on steps $1$ to $j$. \ Then%
\begin{align*}%
p\left(  X\right)  q_{j}%
\left(  X,Y\right)   &  =\Pr\left[  A\right]
\Pr\left[  B_{X}%
|A\right]  \cdot\Pr\left[  B_{Y}|A\right] \\
&  =p\left(  Y\right)  q_{j}%
\left(  Y,X\right)  .
\end{align*}%
Now let $E\left(  X,Y\right)  $\ denote the event that $X\cap Y=S_{X,Y}%
$,\ where $S_{X,Y}$\ is as in Definition \ref{elgood}. \ Also, let $f_{X}%
$\ be
the input obtained from $X$ that has answer bit $0$, and $g_{X}%
$\ be the input
that has answer bit $1$. \ To apply Theorems \ref{ambthm}%
\ and \ref{classadv},
take $\mathcal{A}=\left\{  f_{X}:X\in D_{h,L}%
\right\}  $ and $\mathcal{B}%
=\left\{  g_{X}:X\in D_{h,L}%
\right\}  $. \ Then take $R\left(  f_{X}%
,g_{Y}%
\right)  =w\left(  X,Y\right)  $\ if $E\left(  X,Y\right)  $\ holds, and
$R\left(  f_{X}%
,g_{Y}\right)  =0$\ otherwise. \ Given $f_{X}\in\mathcal{A}%
$\ and $g_{Y}%
\in\mathcal{B}$\ with $R\left(  f_{X},g_{Y}%
\right)  >0$, and
letting $v$ be a vertex such that $f_{X}%
\left(  v\right)  \neq g_{Y}%
\left(
v\right)  $, we must then have either $v\notin X$\ or $v\notin Y$. \ Suppose
the former case; then%
\begin{align*}%
&  \sum_{f_{X^{\ast}}\in\mathcal{A}~:~f_{X^{\ast}}%
\left(  v\right)  \neq
g_{Y}\left(  v\right)  }R\left(  f_{X^{\ast}}%
,g_{Y}\right) \\
&  \leq\sum_{f_{X^{\ast}}\in\mathcal{A}~:~f_{X^{\ast}%
}\left(  v\right)  \neq
g_{Y}\left(  v\right)  }\frac{p\left(  Y\right)  }%
{L}\sum_{j=0}^{L-1}%
q_{j}\left(  Y,X^{\ast}%
\right)  \leq\varepsilon p\left(  Y\right)  ,
\end{align*}%
since $Y$\ is $\varepsilon$-good. \ Thus%
\[
\theta\left(  g_{Y}%
,v\right)  =\frac{\sum_{f_{X^{\ast}}\in\mathcal{A}%
~:~f_{X^{\ast}%
}\left(  v\right)  \neq g_{Y}\left(  v\right)  }R\left(
f_{X^{\ast}}%
,g_{Y}\right)  }{\sum_{f_{X^{\ast}}\in\mathcal{A}}R\left(
f_{X^{\ast}}%
,g_{Y}\right)  }\leq\frac{\varepsilon p\left(  Y\right)
}%
{9p\left(  Y\right)  /10}%
.
\]
Similarly, if $v\notin Y$\ then $\theta\left(  f_{X}%
,v\right)  \leq
10\varepsilon/9$\ by symmetry. \ Hence%
\begin{align*}%
\upsilon_{\min}  &  =\max_{\substack{f_{X}\in\mathcal{A},~g_{Y}%
\in
\mathcal{B},~v:\\R\left(  f_{X},g_{Y}\right)  >0,\\f_{X}%
\left(  v\right)  \neq
g_{Y}\left(  v\right)  }}%
\min\left\{  \theta\left(  f_{X},v\right)
,\theta\left(  g_{Y}%
,v\right)  \right\}  \leq\frac{\varepsilon}{9/10}%
,\\
\upsilon_{\operatorname*{geom}}  &  =\max_{\substack{f_{X}\in\mathcal{A}%
,~g_{Y}\in\mathcal{B},~v:\\R\left(  f_{X},g_{Y}\right)  >0,\\f_{X}%
\left(
v\right)  \neq g_{Y}\left(  v\right)  }}\sqrt{\theta\left(  f_{X}%
,v\right)
\theta\left(  g_{Y},v\right)  }\leq\sqrt{\frac{\varepsilon}{9/10}%
},
\end{align*}
the latter since $\theta\left(  f_{X}%
,v\right)  \leq1$\ and $\theta\left(
g_{Y},v\right)  \leq1$\ for all $f_{X}%
,g_{Y}%
$ and $v$. We now turn to the
general case, in which a snake drawn from $\mathcal{D}%
_{h,L}$\ is
$\varepsilon$\textit{-}%
good\ with probability at least $9/10$. \ Let $G\left(
X\right)  $\ denote the event that $X$ is $\varepsilon$-good. \ Take
$\mathcal{A}%
^{\ast}=\left\{  f_{X}\in\mathcal{A}%
:G\left(  X\right)  \right\}
$ and $\mathcal{B}^{\ast}=\left\{  g_{Y}%
\in\mathcal{B}%
:G\left(  Y\right)
\right\}  $, and\ take $\allowbreak R\left(  f_{X}%
,g_{Y}%
\right)  $\ as before.
\ Then since%
\[
\sum_{X,Y~:~E\left(  X,Y\right)  }%
w\left(  X,Y\right)  \geq\sum_{X}%
\frac{9}{10}%
p\left(  X\right)  \geq\frac{9}{10}%
,
\]
by the union bound we have%
\begin{align*}
&  \sum_{f_{X}%
\in\mathcal{A}^{\ast},~g_{Y}\in\mathcal{B}^{\ast}}R\left(
f_{X}%
,g_{Y}%
\right) \\
&  \geq\sum_{X,Y~:~G\left(  X\right)  \wedge G\left(  Y\right)  \wedge
E\left(  X,Y\right)  }%
w\left(  X,Y\right) \\
&  \quad\quad\quad-\sum_{X~:~\urcorner G\left(  X\right)  }%
p\left(  X\right)
-\sum_{Y~:~\urcorner G\left(  Y\right)  }%
p\left(  Y\right) \\
&  \geq\frac{9}{10}-\frac{1}{10}-\frac{1}{10}%
=\frac{7}{10}.
\end{align*}
So by Lemma \ref{subgraph}%
, there exist subsets $\widetilde{\mathcal{A}%
}\subseteq\mathcal{A}%
^{\ast}$\ and $\widetilde{\mathcal{B}}\subseteq
\mathcal{B}^{\ast}%
$\ such that for all $f_{X}\in\widetilde{\mathcal{A}}$ and
$g_{Y}%
\in\widetilde{\mathcal{B}}$,%
\begin{align*}
\sum_{g_{Y^{\ast}%
}\in\widetilde{\mathcal{B}}}R\left(  f_{X},g_{Y^{\ast}%
}%
\right)   &  \geq\frac{7p\left(  X\right)  }{20},\\
\sum_{f_{X^{\ast}%
}\in\widetilde{\mathcal{A}}}R\left(  f_{X^{\ast}}%
,g_{Y}%
\right)   &  \geq\frac{7p\left(  Y\right)  }{20}.
\end{align*}%
So for all $f_{X},g_{Y}$\ with $R\left(  f_{X},g_{Y}%
\right)  >0$, and all $v$
such that $f_{X}\left(  v\right)  \neq g_{Y}%
\left(  v\right)  $, either
$\theta\left(  f_{X}%
,v\right)  \leq20\varepsilon/7$\ or $\theta\left(
g_{Y}%
,v\right)  \leq20\varepsilon/7$. \ Hence $\upsilon_{\min}%
\leq
20\varepsilon/7$\ and $\upsilon_{\operatorname*{geom}}%
\leq\sqrt{20\varepsilon
/7}$.
\end{proof}
 
\section{Specific Graphs\label{GRAPHS}}

In this section we apply the `snake method' developed in Section \ref{SNAKE}
to specific examples of graphs: the Boolean hypercube in Section
\ref{BOOLEAN}, and the $d$-dimensional cubic grid (for $d\geq3$)\ in Section
\ref{DDIM}.

\subsection{Boolean Hypercube\label{BOOLEAN}}

Abusing notation, we let $\left\{  0,1\right\}  ^{n}$\ denote the
$n$-dimensional Boolean hypercube---that is, the graph whose vertices are
$n$-bit strings,\ with two vertices adjacent if and only if they have Hamming
distance $1$. \ Given a vertex $v\in\left\{  0,1\right\}  ^{n}$, we let
$v\left[  0\right]  ,\ldots,v\left[  n-1\right]  $\ denote the $n$ bits of
$v$, and let $v^{\left(  i\right)  }$ denote the neighbor obtained by flipping
bit $v\left[  i\right]  $. \ In this section we lower-bound
$\operatorname*{RLS}\left(  \left\{  0,1\right\}  ^{n}\right)  $\ and
$\operatorname*{QLS}\left(  \left\{  0,1\right\}  ^{n}\right)  $.

Fix a `snake head' $h\in\left\{  0,1\right\}  ^{n}$ and take $L=2^{n/2}/100$.
\ We define the snake distribution $\mathcal{D}_{h,L}$\ via what we call a
\textit{coordinate loop}, as follows. \ Starting from $x_{0}=h$,\ for each $t$
take $x_{t+1}=x_{t}$\ with $1/2$ probability, and $x_{t+1}=x_{t}^{\left(
t\operatorname{mod}n\right)  }$\ with $1/2$ probability. \ The following is a
basic fact about this distribution. \vspace{-0.1in}

\begin{proposition}
\label{mixtime}The coordinate loop mixes completely in $n$ steps, in the sense
that if $t^{\ast}\geq t+n$, then $x_{t^{\ast}}$\ is a uniform random
vertex\ conditioned on $x_{t}$.
\end{proposition}

\vspace{-0.1in} We could also use the random walk distribution, following
Aldous \cite{aldous}. \ However, not only is the coordinate loop distribution
easier to work with (since it produces fewer self-intersections), it also
yields a better lower bound (since it mixes completely in $n$ steps, as
opposed to approximately in $n\log n$ steps).

We first upper-bound the probability, over $X$, $j$, and $Y\left[  j\right]
$, that $X\cap Y\neq S_{X,Y}$ (where $S_{X,Y}$\ is as in Definition
\ref{elgood}).

\begin{lemma}
\label{intersect}Suppose $X$ is drawn from $\mathcal{D}_{h,L}$, $j$ is drawn
uniformly from $\left\{  0,\ldots,L-1\right\}  $, and $Y\left[  j\right]
$\ is drawn from $\mathcal{D}_{x_{j},j}$. \ Then $\Pr_{X,j,Y\left[  j\right]
}\left[  X\cap Y=S_{X,Y}\right]  \geq0.9999$.
\end{lemma}
\vspace{-0.1in}
\begin{proof}
Call a \textit{disagreement}
a vertex $v$\ such that%
\[
\min\left\{  t:x_{t}%
=v\right\}  \neq\min\left\{  t^{\ast}:y_{t^{\ast}%
}%
=v\right\}  .
\]
Clearly if there are no disagreements then $X\cap Y=S_{X,Y}%
$. \ If $v$\ is a
disagreement, then by the definition of $\mathcal{D}%
_{h,L}$\ we cannot have
both $t>j-n$\ and $t^{\ast}%
>j-n$. \ So by Proposition \ref{mixtime}, either
$y_{t^{\ast}}%
$\ is uniformly random conditioned on $X$, or $x_{t}%
$\ is
uniformly random conditioned on $Y\left[  j\right]  $. \ Hence $\Pr
_{X,j,Y\left[  j\right]  }%
\left[  x_{t}=y_{t^{\ast}}\right]  =1/2^{n}%
$. \ So
by the union bound,%
\[
\Pr_{X,j,Y\left[  j\right]  }%
\left[  X\cap Y\neq S_{X,Y}\right]
\leq\frac{L^{2}}{2^{n}}%
=0.0001.
\]
\end{proof}
 \vspace{-0.1in} We now argue that, unless $X$
spends a `pathological' amount of time in one part of the hypercube, the
probability of any vertex $v$ being hit when $X$ flicks its tail is small.
\ To prove this, we define a notion of \textit{sparseness}, and then show that
(1) almost all snakes drawn from $\mathcal{D}_{h,L}$\ are sparse (Lemma
\ref{hammingball}), and (2) sparse snakes are unlikely to hit any given vertex
$v$ (Lemma \ref{sparsegood}). \vspace{-0.1in}

\begin{definition}
\label{sparse}Given vertices $v,w$ and $i\in\left\{  0,\ldots,n-1\right\}  $,
let $\Delta\left(  x,v,i\right)  $\ be the number of steps needed to reach $v$
from $x$ by first setting $x\left[  i\right]  :=v\left[  i\right]  $, then
setting $x\left[  i-1\right]  :=v\left[  i-1\right]  $, and so on. \ (After we
set $x\left[  0\right]  $\ we wrap around to $x\left[  n-1\right]  $.) \ Then
$X$ is \textit{sparse} if there exists a constant $c$ such that for all
$v\in\left\{  0,1\right\}  ^{n}$ and all $k$,%
\[
\left\vert \left\{  t:\Delta\left(  x_{t},v,t\operatorname{mod}n\right)
=k\right\}  \right\vert \leq cn\left(  n+\frac{L}{2^{n-k}}\right)  .
\]
\end{definition}
\begin{lemma}
\label{hammingball}If $X$ is drawn from $\mathcal{D}_{h,L}$, then $X$ is
sparse with probability $1-o\left(  1\right)  $.
\end{lemma}
\vspace{-0.2in}
\begin{proof}%
For each $i\in\left\{  0,\ldots,n-1\right\}  $, the number of $t\in\left\{
0,\ldots,L-1\right\}  $\ such that $t\equiv i\left(  \operatorname{mod}%
n\right)  $\ is at most $L/n$.\ \ For such a $t$, let $E_{t}%
^{\left(
v,i,k\right)  }$\ be the event that $\Delta\left(  x_{t}%
,v,i\right)  \leq k$;
then $E_{t}^{\left(  v,i,k\right)  }%
$\ holds if and only if%
\[
x_{t}%
\left[  i\right]  =v\left[  i\right]  ,\ldots,x_{t}%
\left[  i-k+1\right]
=v\left[  i-k+1\right]
\]
(where we wrap around to $x_{t}%
\left[  n-1\right]  $\ after reaching
$x_{t}%
\left[  0\right]  $). \ This occurs with probability $2^{k}/2^{n}%
$\ over
$X$. \ Furthermore, by Proposition \ref{mixtime}, the $E_{t}%
^{\left(
v,i,k\right)  }%
$\ events for different $t$'s are independent. \ So let%
\[
\mu_{k}%
=\frac{L}{n}\cdot\frac{2^{k}}{2^{n}}%
;
\]
then for fixed $v,i,k$, the expected number of $t$'s\ for which $E_{t}%
^{\left(  v,i,k\right)  }$\ holds is at most $\mu_{k}%
$. \ Thus by a Chernoff
bound, if $\mu_{k}\geq1$\ then%
\[
\Pr_{X}%
\left[  \left|  \left\{  t:E_{t}^{\left(  v,i,k\right)  }%
\right\}
\right|  >cn\cdot\mu_{k}\right]  <\left(  \frac{e^{cn-1}%
}{\left(  cn\right)
^{cn}}\right)  ^{\mu_{k}}<\frac{1}{2^{2n}}%
\]
for sufficiently large $c$. \ Similarly, if $\mu_{k}%
<1$\ then%
\[
\Pr_{X}\left[  \left|  \left\{  t:E_{t}%
^{\left(  v,i,k\right)  }%
\right\}
\right|  >cn\right]  <\left(  \frac{e^{cn/\mu_{k}-1}}%
{\left(  cn/\mu
_{k}\right)  ^{cn/\mu_{k}}}\right)  ^{\mu_{k}}<\frac{1}%
{2^{2n}}%
\]
for sufficiently large $c$. \ By the union bound, then,%
\[
\left|  \left\{  t:E_{t}%
^{\left(  v,i,k\right)  }\right\}  \right|  \leq
cn\cdot\left(  1+\mu_{k}%
\right)  =c\left(  n+\frac{L}{2^{n-k}}%
\right)
\]
for every $v,i,k$ triple \textit{simultaneously}
with probability at least
$1-n^{2}2^{n}/2^{2n}%
=1-o\left(  1\right)  $. \ Summing over all $i$'s produces
the additional factor of $n$.
\end{proof}
\begin{lemma}
\label{sparsegood}If $X$ is sparse, then for every $v\in\left\{  0,1\right\}
^{n}$,%
\[
\Pr_{j,Y}\left[  v\in Y\left[  j\right]  \right]  =O\left(  \frac{n^{2}}%
{L}\right)  .
\]
\end{lemma}
\vspace{-0.1in}
\begin{proof}%
By assumption, for every $k\in\left\{  0,\ldots,n\right\}  $,%
\begin{align*}%
\Pr_{j}\left[  \Delta\left(  x_{j},v,j\operatorname{mod}%
n\right)  =k\right]
&  \leq\frac{\left|  \left\{  t:\Delta\left(  x_{t}%
,v,t\operatorname{mod}%
n\right)  =k\right\}  \right|  }{L}%
\\
&  \leq\frac{cn}{L}\left(  n+\frac{L}{2^{n-k}}\right)  .
\end{align*}%
Consider the probability that $v\in Y\left[  j\right]  $\ in the event that
$\Delta\left(  x_{j}%
,v,j\operatorname{mod}n\right)  =k$.\ \ Clearly%
\[
\Pr_{Y}%
\left[  v\in\left\{  y_{j-n+1},\ldots,y_{j}\right\}  \right]
=\frac{1}{2^{k}%
}.
\]
Also, Proposition \ref{mixtime}%
\ implies that for every $t\leq j-n$, the
probability that $y_{t}%
=v$\ is $2^{-n}$. \ So by the union bound,%
\[
\Pr_{Y}%
\left[  v\in\left\{  y_{0},\ldots,y_{j-n}\right\}  \right]
\leq\frac{L}%
{2^{n}}.
\]
Then $\Pr_{j,Y}%
\left[  v\in Y\left[  j\right]  \right]  $\ equals%
\begin{align*}%
&  \sum_{k=0}^{n}\left(
\begin{array}
[c]{c}%
\Pr_{j}%
\left[  \Delta\left(  x_{j},v,j\operatorname{mod}%
n\right)  =k\right]
\cdot\\
\Pr_{Y}%
\left[  v\in Y\left[  j\right]  ~|~\Delta\left(  x_{j}%
,v,j\operatorname{mod}n\right)  =k\right]
\end{array}%
\right) \\
&  \leq\sum_{k=0}^{n}\frac{cn}{L}\left(  n+\frac{L}{2^{n-k}%
}\right)  \left(
\frac{1}{2^{k}}+\frac{L}{2^{n}}%
\right)  =O\left(  \frac{cn^{2}}{L}\right)
\end{align*}%
as can be verified by breaking the sum into cases and doing some manipulations.
\end{proof}%
\vspace{-0.1in}
The main result follows easily:
\vspace{-0.1in}
\begin{theorem}
\label{boolean}%
\[
\operatorname*{RLS}\left(  \left\{  0,1\right\}  ^{n}\right)  =\Omega\left(
\frac{2^{n/2}}{n^{2}}\right)  ,~~\operatorname*{QLS}\left(  \left\{
0,1\right\}  ^{n}\right)  =\Omega\left(  \frac{2^{n/4}}{n}\right)  .
\]
\end{theorem}
\vspace{-0.1in}
\begin{proof}
Take $\varepsilon=n^{2}/2^{n/2}%
$. \ Then by Theorem \ref{kappathm}%
,\ it
suffices to show that a snake $X$ drawn from $\mathcal{D}%
_{h,L}%
$\ is $O\left(
\varepsilon\right)  $-good with probability at least\ $9/10$. \ First, since%
\[
\Pr_{X,j,Y\left[  j\right]  }%
\left[  X\cap Y=S_{X,Y}\right]  \geq0.9999
\]
by Lemma \ref{intersect}%
, Markov's inequality shows that%
\[
\Pr_{X}%
\left[  \Pr_{j,Y\left[  j\right]  }\left[  X\cap Y=S_{X,Y}%
\right]
\geq\frac{9}{10}\right]  \geq\frac{19}{20}%
.
\]
Second, by Lemma \ref{hammingball}%
, $X$ is sparse with probability $1-o\left(
1\right)  $, and by Lemma \ref{sparsegood}%
, if $X$ is sparse then%
\[
\Pr_{j,Y}%
\left[  v\in Y\left[  j\right]  \right]  =O\left(  \frac{n^{2}}%
{L}%
\right)  =O\left(  \varepsilon\right)
\]
for every $v$.\ \ So both requirements of Definition \ref{elgood}%
\ hold
simultaneously with probability at least $9/10$.
\end{proof}
 
\subsection{Constant-Dimensional Grid Graph\label{DDIM}}

In the Boolean hypercube case, we defined $\mathcal{D}_{h,L}$\ by a
`coordinate loop' instead of the usual random walk mainly for convenience.
\ When we move to the $d$-dimensional grid, though, the drawbacks of random
walks become more serious: first, the mixing time is too long, and second,
there are too many self-intersections, particularly if $d\leq4$. \ Our snake
distribution will instead use straight lines of randomly chosen lengths
attached at the endpoints, as in Figure 2.%
\begin{figure}
[ptb]
\begin{center}
\includegraphics[
trim=4.168388in 5.539468in 4.183915in 0.560694in,
height=2.015in,
width=2.4267in
]%
{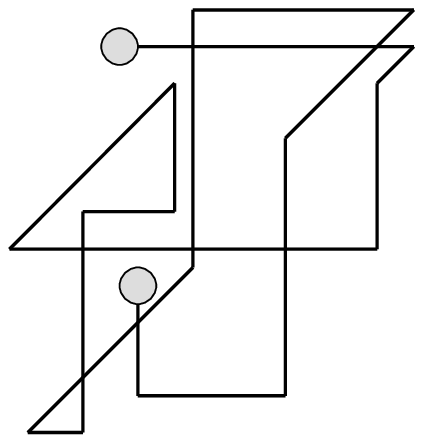}%
\caption{In $d=3$ dimensions, a snake drawn from $\mathcal{D}_{h,L}$\ moves a
random distance left or right, then a random distance up or down, then a
random distance inward or outward, etc.}%
\label{linefig}%
\end{center}
\end{figure}
Let $G_{d,N}$\ be a $d$-dimensional grid graph with $d\geq3$. \ That is,
$G_{d,N}$\ has $N$ vertices of the form $v=\left(  v\left[  0\right]
,\ldots,v\left[  d-1\right]  \right)  $, where each $v\left[  i\right]  $\ is
in $\left\{  1,\ldots,N^{1/d}\right\}  $ (we assume for simplicity that $N$ is
a $d^{th}$\ power). \ Vertices $v$ and $w$ are adjacent if and only if
$\left\vert v\left[  i\right]  -w\left[  i\right]  \right\vert =1$\ for some
$i\in\left\{  0,\ldots,d-1\right\}  $, and $v\left[  j\right]  =w\left[
j\right]  $\ for all $j\neq i$ (so $G_{d,N}$\ does not wrap around at the boundaries).

We take $L=\sqrt{N}/100$, and define the snake distribution $\mathcal{D}%
_{h,L}$\ as follows. \ Starting from $x_{0}=h$, for each $T$\ we take
$x_{N^{1/d}\left(  T+1\right)  }$\ identical to $x_{N^{1/d}T}$, but with the
$\left(  T\operatorname{mod}d\right)  ^{th}$\ coordinate $x_{N^{1/d}\left(
T+1\right)  }\left[  T\operatorname{mod}d\right]  $\ replaced by a uniform
random value in $\left\{  1,\ldots,N^{1/d}\right\}  $. \ We then take the
vertices $x_{N^{1/d}T+1},\ldots,x_{N^{1/d}T+N^{1/d}-1}$\ to lie along the
shortest path from $x_{N^{1/d}T}$\ to $x_{N^{1/d}\left(  T+1\right)  }%
$,\ `stalling'\ at $x_{N^{1/d}\left(  T+1\right)  }$\ once that vertex has
been reached. \ We call%
\[
\Phi_{T}=\left(  x_{N^{1/d}T},\ldots,x_{N^{1/d}T+N^{1/d}-1}\right)
\]
a \textit{line} of vertices, whose \textit{direction} is $T\operatorname{mod}%
d$. \ As in the Boolean hypercube case, we have: \vspace{-0.1in}

\begin{proposition}
\label{mixtime2}$\mathcal{D}_{h,L}$ mixes completely in $dN^{1/d}$ steps, in
the sense that if $T^{\ast}\geq T+d$, then $x_{N^{1/d}T^{\ast}}$\ is a uniform
random vertex\ conditioned on $x_{N^{1/d}T}$.
\end{proposition}
\vspace{-0.1in} Lemma \ref{intersect}\ in Section \ref{BOOLEAN}\ goes through
essentially without change. \vspace{-0.1in}
\begin{definition}
\label{sparse2}Letting $\Delta\left(  x,v,i\right)  $\ be as before, we say
$X$ is \textit{sparse} if there exists a constant $c$ (possibly dependent on
$d$) such that for all vertices $v$ and all $k$,%
\begin{align*}
&  \left|  \left\{  t:\Delta\left(  x_{t},v,\left\lfloor t/N^{1/d}%
\right\rfloor \operatorname{mod}d\right)  =k\right\}  \right| \\
&  \leq\left(  c\log N\right)  \left(  N^{1/d}+\frac{L}{N^{1-k/d}}\right)  .
\end{align*}
\end{definition}
\begin{lemma}
\label{hammingball2}If $X$ is drawn from $\mathcal{D}_{h,L}$, then $X$ is
sparse with probability $1-o\left(  1\right)  $.
\end{lemma}
\vspace{-0.2in}
\begin{lemma}
\label{sparsegood2}If $X$ is sparse, then for every $v\in G_{d,N}$,%
\[
\Pr_{j,Y}\left[  v\in Y\left[  j\right]  \right]  =O\left(  \frac{N^{1/d}\log
N}{L}\right)  ,
\]
where the big-$O$ hides a constant dependent on $d$.
\end{lemma}
\vspace{-0.1in}
The proofs of Lemmas \ref{hammingball2}\ and \ref{sparsegood2}\ are omitted
from this abstract, since they involve no new ideas beyond those of Lemmas
\ref{hammingball} and \ref{sparsegood}. \ Taking $\varepsilon=\left(  \log
N\right)  /N^{1/2-1/d}$\ we get, by the same proof as for Theorem
\ref{boolean}:
\vspace{-0.1in}
\begin{theorem}
\label{grid}Neglecting a constant dependent on $d$, for all $d\geq3$%
\begin{align*}
\operatorname*{RLS}\left(  G_{d,N}\right)   &  =\Omega\left(
\frac{N^{1/2-1/d}}{\log N}\right)  ,\\
\operatorname*{QLS}\left(  G_{d,N}\right)   &  =\Omega\left(  \sqrt
{\frac{N^{1/2-1/d}}{\log N}}\right)  .
\end{align*}
\end{theorem}

\section{Acknowledgments}

I thank Andris Ambainis for suggesting an improvement to Proposition
\ref{upper}; David Aldous, Christos Papadimitriou, Yuval Peres, and Umesh
Vazirani for discussions during the early stages of this work; and Ronald de
Wolf and the anonymous reviewers for helpful comments.

\end{document}